\documentclass[11pt,a4paper]{article} 
\usepackage{jheppub}

\usepackage{slashed}
\usepackage{youngtab}

\newcommand{\beq}{\begin{equation}}
\newcommand{\eeq}{\end{equation}}

\newcommand{\beqa}{\begin{eqnarray}}
\newcommand{\eeqa}{\end{eqnarray}}

\newcommand{\bn}{\begin{equation}}
\newcommand{\en}{\end{equation}}
\newcommand{\by}{\begin{eqnarray}}
\newcommand{\ey}{\end{eqnarray}}
\newcommand{\al}{\alpha}
\newcommand{\be}{\beta}
\newcommand{\ga}{\gamma}

\newcommand{\de}{\delta}
\newcommand{\ep}{\epsilon}
\newcommand{\om}{\omega}
\newcommand{\Om}{\Omega}
\newcommand{\si}{\sigma}
\newcommand{\Lam}{\Lambda}
\newcommand{\tLam}{\tilde \Lambda}
\newcommand{\thalf}{\tfrac{1}{2}}
\newcommand{\pd}{\partial}
\newcommand{\te}{\tilde e}
\newcommand{\tf}{\tilde f}
\newcommand{\tF}{\tilde F}
\newcommand{\tP}{\tilde P}
\newcommand{\tom}{\tilde \om}
\newcommand{\tOm}{\tilde \Om}
\newcommand{\tM}{\tilde M}
\newcommand{\tb}{\tilde b}
\newcommand{\tD}{\tilde D}
\newcommand{\tK}{\tilde K}
\newcommand{\we}{\wedge}

\newcommand\fverb{\setbox\fverbbox=\hbox\bgroup\verb}
\newcommand\fverbdo{\egroup\medskip\noindent%
            \fbox{\unhbox\fverbbox}\ }
\newcommand\fverbit{\egroup\item[\fbox{\unhbox\fverbbox}]}
\newbox\fverbbox

\newcommand{\nablaslash}{\not{\hbox{\kern-3pt $\nabla$}}}

\title{Towards an exact frame formulation of conformal higher spins in three dimensions}

\author{Bengt E.W.~Nilsson
\\
Fundamental Physics\\
Chalmers University of Technology\\
SE-412 96 G\"oteborg, Sweden\\

{\tt {\footnotesize  tfebn@chalmers.se}}}

\abstract{ In this paper we discuss some aspects of the frame formulation of 
conformal higher spins in three dimensions. We give some exact formulae for the
coupled spin two - spin three part of the full higher spin theory and propose a star product Lagrangian
for all spins from two and up. Since there is no consistent Lagrangian formulation based on the Poisson bracket
 we start the construction from the field equations in this approximation of the star product.
 The higher spin algebra
is then realized in terms of classical variables which leads to certain important simplifications
that we take advantage of. The suggested structure of the all-spin Lagrangian given here
is, however,  obtained using an expansion of the star product beyond the Poisson bracket in terms of multi-commutators and the Lagrangian
should be viewed as a starting point for the derivation of the full theory based on a star product. How to do this
is explained as well as how to include the
coupling to scalar fields. We also comment on the AdS/CFT relation to four dimensions.}
 

\toccontinuoustrue
\begin{document}
\maketitle



\section{Introduction}


Conformal symmetries in three dimensions play an important role in string theory and  many condensed matter systems.
Through the AdS/CFT correspondence we now believe that the M2-brane is described by a conformal matter theory that 
is  coupled to a spin one Chern-Simons theory. In the condensed matter context much recent work has been devoted to
studying strongly coupled phenomena by an $AdS/CFT$ mapping of the system to an AdS gravitational theory which is weakly coupled and lives 
in one dimension higher then the conformal system in question.
These latter considerations involve mostly  conformal vector models which, if exhibiting higher spin symmetries and conserved currents, are  mapped to Vasiliev-type higher spin (HS) theories
on the gravity side, see for instance \cite{Vasiliev:2012vf} which also contains results directly relevant for the discussion below. Particularly for $AdS_4/CFT_3$ this has resulted in intriguing speculations about relations between bosonic and fermionic
matter theories in three dimensions \cite{Sundborg:2000wp,Sezgin:2002rt, Klebanov:2002ja,Sezgin:2003pt} with features reminiscent of 
bosonization \cite{Aharony:2012nh,Chang:2012kt}\footnote{A related connection between the $A$ and $B$ versions of Vasiliev's theory in $AdS_4$
was observed by Leigh and Petkou in \cite{Leigh:2003gk}. Deformation by double trace operators for all spins needed for the transition between the $A$ and $B$ versions were introduced in \cite{Sezgin:2002rt}, however, for other reasons than changing boundary conditions.}. 
 
Fields of all spins in Vasiliev-type $AdS$ higher spin theories must be given specific boundary conditions in order to completely define the theory.
This can be done using either Dirichlet or Neumann boundary conditions, or conditions that interpolate between them.
For spin zero this was clear already in the early investigations of Breitenlohner and Freedman who found the ranges of masses where these could be implemented.
The spin one case was discussed by Witten in \cite{Witten:2003ya} while for spin two the situation is a bit more subtle. However,  a number of similar results have been obtained 
by Petkou and Leigh \cite{Leigh:2003ez}, see also \cite{Leigh:2007wf, deHaro:2008gp}
   and references therein.
For spin three and higher recent results by Giombi et al \cite{Giombi:2013yva} (see also \cite{Giombi:2013fka, Tseytlin:2013jya,Tseytlin:2013fca})  indicate that Neumann boundary conditions are possible and even crucial in certain considerations. The importance of Neumann boundary conditions was also noted in \cite{Nilsson:2012ky}.
The conformal duals of AdS gauge fields with Neumann boundary conditions contain gauged symmetries which play a central role in recent attempts to relate higher spin theories to string/M theory and ABJM/ABJ theories, as discussed in, e.g., \cite{Chang:2012kt}. This reasoning seems to indicate that conformal higher spin systems may be important in further studies of these connections. Such systems in three dimensions are special since they are based on Chern-Simons-type constructions and do not contain any degrees of freedom. It would be interesting to see if such higher spin theories
can be coupled to matter systems that we normally connect to M2-branes, e.g., ABJM/ABJ and BLG theories. In fact, the spin two version of such constructions are available 
with some surprising features like an $O(N)$ vector-like generalization of the BLG model and a number of non-trivial background solutions appearing as a consequence of a new 
set of potential terms \cite{Chu:2009gi,Nilsson:2013fya}, see also \cite{Gran:2008qx,Chu:2010fk,Gran:2012mg}.

Here we study  the coupled spin 2 - spin 3 conformal system in three dimensions and obtain an exact formulation thereof\footnote{In this brief paper only some equations are given in full detail. A complete account  will be given elsewhere.}. This theory is a higher derivative Chern-Simons type theory
generalizing the well-known spin 2 case   \cite{Pope:1989vj,Horne:1988jf}. For spin above two there are Vasiliev-type higher spin (star product) constructions of many interacting theories in $AdS$ spaces but these methods seem hard to use in the conformal context, see for instance \cite{Vasiliev:2012vf}\footnote{Only recently have explicit amplitudes in $AdS$ beyond three-point couplings been obtained
 in \cite{Didenko:2012tv} using a method invented in \cite{Colombo:2012jx}.}. Thus, we can not exploit these techniques in our case to obtain  explicit
expressions  in terms of objects like curvatures etc. 

In this paper  we obtain a formulation which is 
exact in the sense that all higher point interaction terms are kept intact and not truncated to only three or four point. There is, however, another  simplifying 
feature build into our 
construction in that star products are initially replaced by Poisson brackets. This can be viewed as a certain "classical limit" of the higher spin algebra based on the star product. The Poisson version of the three dimensional conformal  higher spin theory was studied at the linear level by Blencowe \cite{Blencowe:1988gj} and by Pope and Townsend \cite{Pope:1989vj}. The Chern-Simons method used in these papers was soon afterwards studied in the spin two case by Horne and Witten in \cite{Horne:1988jf} who  
 analyzed this in full detail and gave a proper interpretation of symmetries etc. The star product version of the theory was discussed at the linear level in many 
papers, see, e.g., \cite{Shaynkman:2001ip} (see also  \cite{Fradkin:1989xt,Fradkin:1989yd,Vasiliev:1992ix,Segal:2002gd}).  The coupling to matter fields, i.e., scalar and spinor fields, is intricate and requires unfolding as discussed at the linear level in, e.g.,  \cite{Vasiliev:2012vf}. The complications of taking these constructions to the non-linear level is also mentioned there. 

Here we advocate a different approach
for obtaining the full non-linear structure of the conformal higher spin theory and its coupling to scalars. As explained below one can start from a solution of the zero field strength equations of the higher spin theory based on the Poisson bracket, construct the Lagrangian and then step by step introduce the star product and the couplings to scalar fields.  This approach makes use of fully covariant (under the usual spin 2 symmetries  in three dimensions) tensor equations at all stages.

\section{The approach}

The starting point for our exact treatment of the three-dimensional coupled conformal spin 2 - spin 3  system is the Poisson bracket realization of the higher spin algebra.
We will  closely  follow the presentation of Pope and Townsend in \cite{Pope:1989vj}. The higher spin algebra contains generators $G(a,b)=q^{\al_1}...q^{\al_a}p_{\be_1}...p_{\be_b}$ where the classical "phase space" variables $q^{\al}$ and $p_{\be}$ are bosonic 
$SO(1,2)$ spinors satisfying $\{q^{\al}, p_{\be}\}_{PB}=\de^{\al}_{\be}$.
This leads to the grading property $\{G(a,b),G(a',b')\}_{PB}=G(a+a'-1,b+b'-1)$ where the right  hand side would in a quantized version, using a star product or other, of the algebra contain additional terms with lower grade (spin). The generators of the "classical" higher spin algebra are then gauged with the  spin content of the corresponding gauge 
fields (generalizing the dreibein) given by $s=\thalf(a+b)+1$ since the spin  of the generator itself is $\thalf(a+b)$ and the gauge field has an extra vector index. Denoting the generators  $G(a,b)$ with a given spin collectively as $G(\thalf(a+b))$, this means for instance that the generators related to  fields with spin 2 and spin 3, namely  $G(1)$ and $G(2)$, satisfy $\{G(1),G(1)\}_{PB}=G(1)$ and  $\{G(1),G(2)\}_{PB}=G(2)$ which provide the starting-point for our discussion below.
Brackets like $\{G(2),G(2)\}_{PB}=G(3)$ and higher ones will only enter the discussion towards the end of the paper when specific terms in the all-spin Lagrangian are discussed. 
 
 At this point we change the notation for the bracket and denote the Poisson bracket by $[..,..]$ and use $[..,..]_*$ for the full structure of the star product. Of course, as mentioned above, quantizing the variables $q,p$ will result in further terms in the algebra coming from multiple commutators of the operators $\hat q,\hat p$ leading to new terms with lower spin on the right hand sides. This, in fact, will be the language used towards the end of the paper when discussing the Lagrangian for $all$ spins.
 
 The aim of this paper is to follow the strategy laid out  in \cite{Pope:1989vj}, that is, gauge the higher spin  algebra by introducing a one-form connection $A$ valued in the algebra
 based on the Poisson bracket and solve the corresponding zero field strength conditions.  These $F(A)=0$ equations  may be viewed as the field equations coming from a Lagrangian in the form of a Chern-Simons term $S[A]=\thalf\int(AdA+\tfrac{2}{3}A^3)$ invariant under the gauge transformations $\de A=D\Lam=d\Lam+[A,\Lam]$ corresponding to the higher spin version of the conformal group  $SO(2,3)$. The action we present below resembles this but is, however, given in terms of only the various spin connections and curvature tensors for
 the higher spin frame fields generalizing the dreibein $e_{\mu}{}^a$, namely $e_{\mu}{}^{a_1....a_n}$ for each spin $s=n+1$ sector of the full higher spin theory. In other words, the 
 final result is here written in terms of the independent fields $e_{\mu}{}^{a_1....a_n}$ obtained by solving the algebraic subset of equations in  $F(A)=0$. The aim here is to be brief saving the details for another publication \cite{Nilsson:2015AB}.
 
For the present  purposes we need the terms in $A$  with $a+b=2$
 \beq
 A(1)=e_aP^a+\om_aM^a+bD+f_aK^a,
 \eeq
 and  $a+b=4$
 \beq
 A(2)=e_{ab}P^{ab}+\te_{ab}\tP^{ab}+\te_a\tP^a+\tom_{ab}\tM^{ab}+\tom_{a}\tM^{a}+\tb\tD+\tf_{ab}\tK^{ab}+\tf_{a}\tK^{a}+f_{ab}K^{ab},
 \eeq
 where all representations are  irreducible written in terms of symmetrized and traceless sets of flat $SO(1,2)$ vector indices $ab...$.
Here the generators are denoted as $P,M,D,K$ (with or without a tilde (see below)) and similarly for the corresponding gauge fields $e,\om,b,f$.
The indices on the generators follow from their explicit form in terms of the classical variables $q^{\al}$ and $p_{\al}$: 
\beq
G^{a_1....a_n,b,\, c_1...c_m}=(-\thalf\si^{a_1})_{\al_1\al_2}q^{\al_1}q^{\al_2}....(-\thalf\si^{b})_{\al_{2n+1}}{}^{\ga_{2m+1}}q^{\al_{2n+1}}p_{\ga_{2m+1}}
(-\thalf\si^{c_1})^{\ga_1\ga_2}p_{\ga_1}p_{\ga_2}..,
\eeq 
where the $(\si^{b})$ factor is replaced by a delta leading to a contracted pair $q^{\al}p_{\al}$ in some generators, or is just  absent if the number of $q^{\al}$s (and $p_{\al}$s) is even.
 For instance, we have defined
$\tP^{ab}(3,1)=\tfrac{1}{4}(\si^{(a})_{\al_1\al_2}q^{\al_1}q^{\al_2}(\si^{b)})_{\al_3}{}^{\ga}q^{\al_3}p_{\ga}$ while
$\tP^{a}(3,1)=\tfrac{1}{4}(\si^{a})_{\al_1\al_2}q^{\al_1}q^{\al_2}q^{\al}p_{\al}$. Note that $\tP^{ab}(3,1)$ is defined to be symmetric but that the tracelessness
is due to a Fierz identity. These rules lead to the following  content of irreps for the generators: for spin 2
\beq
P^a(2,0), M^a(1,1), D(1,1), K^a(0,2),
\eeq
 and  for spin 3 
\beq
P^{ab}(4,0), \tP^{ab}(3,1), \tP^a(3,1), \tM^{ab}(2,2), \tM^a(2,2), \tD(2,2), \tK^{ab}(1,3), \tK^a(3,1), K^{ab}(0,4).
\eeq

As a warm-up we start with  the spin two case, following \cite{Pope:1989vj} (or  \cite{Horne:1988jf}), and solve the zero field strength conditions $F=0$ in a step by step manner as follows. The relevant equations are
(where superscripts L(orentz) and D(ilatation) refer to the vector and scalar parts of the $(1,1)$ sector)
\beq
F_a(2,0)=de_a(2,0)+\{\om(1,1),e(2,0)\}|_{P^a}+\{b(1,1),e(2,0)\}|_{P^a}=0,
\eeq
\beq
F_a^L(1,1)=d\om_a(1,1)+\thalf\{\om(1,1),\om(1,1)\}|_{M^a}+\{e(2,0),f(0,2)\}|_{M^a}=0,
\eeq
\beq
F^D(1,1)=db(1,1)+\{e(2,0),f(0,2)\}|_D=0,
\eeq
\beq
F_a(0,2)=df_a(0,2)+\{\om(1,1),f(0,2)\}|_{K^a}+\{b(1,1),f(0,2)\}|_{K^a}=0,
\eeq
where $\{\om(1,1),e(2,0)\}|_{P^a}=\om_b\wedge e_c[M^b,P^c]|_{P^a}=\om_b\wedge e_c\ep^{bc}{}_aP^a|_{P^a}=\om_b\wedge e_c\ep^{bc}{}_a$ and similarly for the other terms. 
The symmetry parameters are
\beq
\Lam_a(2,0), \Lam_a^L(1,1),\Lam^D(1,1), \Lam_a(0,2).
\eeq
Note that from the explicit form of the transformation rules (see \cite{Pope:1989vj}) it follows that we can use $\Lam_a(0,2)$ to set the dilatation gauge field to zero: $b_{\mu}=0$. The remaining parameters then correspond to the symmetries of conformal systems, namely diffeomorphisms, Lorentz and dilatations. The exact relation is, however, non-trivial and only possible to establish on-shell
  \cite{Horne:1988jf}. The equations above then read, with $D=d+\om(1,1)$, $R(1,1)=d\om(1,1)+\om(1,1)\wedge \om(1,1)$, 
\beq
F_a(2,0)=De_a(2,0)=0,
\eeq
\beq
F_a^L(1,1)=R_a(1,1)+\{e(2,0),f(0,2)\}|_{M^a}=0,
\eeq
\beq
F^D(1,1)=\{e(2,0),f(0,2)\}|_D=0,
\eeq
\beq
F_a(0,2)=Df_a(0,2)=0.
\eeq
The first three of these equations are easily solved. The first one is the zero torsion equation giving the spin connection in terms of the dreibein. The second one
can be solved with similar methods with the result that $f_{\mu}{}^a(0,2)$ is just one half of the spin 2 Schouten tensor which is constrained to be symmetric as a consequence of the third equation. Thus,  the relation to the Schouten tensor is \cite{Pope:1989vj}
\beq
f_{\mu\nu}(0,2)=\thalf(R_{\mu\nu}-\tfrac{1}{4}g_{\mu\nu}R)
\eeq
and the constraint is thus automatically satisfied. In the spin 3 case a number of much more involved constraints appear and it will take some effort to analyze them.
The last equation in the above list is then just the spin 2 Cotton equation which is third order in derivatives. We will demonstrate how to find the corresponding exact result for spin 3 below, then in terms of a fifth order field equation. In order to see how this works we now repeat the steps in the spin 2 case for the spin 3 part of the zero field strength equations. 

\section{Spin three}

We now turn to the spin three part of the zero field strength equations and the transformation rules for each field. We will utilize these equations in a step by step procedure that hopefully will shed some light on how to repeat this for the higher spins. Looking briefly at spin four and five indicates that the main structure to be spelt out in this paper is the same for all spins but the route to get to the answer  may vary somewhat between different spins. Note that the analysis in this section is in terms of Poisson brackets. The corresponding star bracket results will be discussed in the following section. This section thus completes the results in \cite{Pope:1989vj} for spin 3 by obtaining the full spin 2 covariant solution as well as providing a full
analysis of the constraints. In \cite{Pope:1989vj} it was shown by counting representations at the linear level that the equation $F=0$ contains enough information to relate all fields to the spin $n+1$ frame fields 
$e_{\mu}{}^{a_1...a_n}$ but, for instance,  the role of the constraints were not discussed.

Using the spin 2 Lorentz covariant derivative $D=d+\om(1,1)$ 
we have
(products of 1-forms in the curly brackets (anti-commutators) below are with wedge products)
\beq
\de e(4,0)=D\Lam(4,0)+[b(1,1),\Lam(4,0)]+[e(2,0),\Lam(3,1)]+[e(3,1),\Lam(2,0)]+[e(4,0),\Lam(1,1)],
\eeq
\beq
F(4,0)=D e(4,0)+\{b(1,1),e(4,0)\}+\{e(2,0),e(3,1)\}=0,
\eeq
\beqa
\de e(3,1)&=&D\Lam(3,1)+[b(1,1),\Lam(3,1)]+[e(2,0),\Lam(2,2)]+[\om(2,2),\Lam(2,0)]\cr
&&+[e(3,1),\Lam(1,1)]+[f(0,2),\Lam(4,0)]+[e(4,0),\Lam(0,2)],
\eeqa
\beq
F(3,1)=D e(3,1)+\{b(1,1),e(3,1)\}+\{e(2,0),\om(2,2)\}+\{f(0,2),e(4,0)\}=0,
\eeq
\beqa
\de \omega(2,2)&=&D\Lam(2,2)+[b(1,1),\Lam(2,2)]+[e(2,0),\Lam(1,3)]+[f(1,3),\Lam(2,0)]\cr
&&+[\omega(2,2),\Lam(1,1)]+[f(0,2),\Lam(3,1)]+[e(3,1),\Lam(0,2)],
\eeqa
\beq
F(2,2)=D \om(2,2)+\{e(2,0),f(1,3)\}+\{f(0,2),e(3,1)\}=0,
\eeq
\beqa
\de f(1,3)&=&D\Lam(1,3)+[e(2,0),\Lam(0,4)]+[f(0,4),\Lam(2,0)]\cr
&&+[f(1,3),\Lam(1,1)]+[f(0,2),\Lam(2,2)]+[\om(2,2),\Lam(0,2)],,
\eeqa
\beq
F(1,3)=D e(1,3)+\{e(2,0),f(0,4)\}+\{f(0,2),\omega(2,2)\}=0,
\eeq
\beq
\de f(0,4)=D\Lam(0,4)+[f(0,4),\Lam(1,1)]+[f(0,2),\Lam(1,3)]+[f(1,3),\Lam(0,2)],
\eeq
\beq
F(0,4)=D f(0,4)+\{f(0,2),f(1,3)\}=0.
\eeq
The first of these equations gives the transformation rules for the independent spin 3 "dreibein" field while the last equation
provides its fifth order spin 3 Cotton-type equation. Apart from setting the field $b(1,1)$ to zero as we saw above in the spin 2 warm-up, the first equation giving the transformation rule
for the spin 3 frame field $e_{\mu}{}^{ab}$ will not be changed. It therefore already at this point indicates which symmetries will be kept intact during the analysis below. These will be discussed further after we have completed the analysis of the whole set of equations.
The exercise is now to use the other equations in a way that
simplifies the description of the spin-3/spin-2 system and provides a formulation that might help generalizing it to even higher spins.

The main objective is thus to express the field $f(0,4)$ explicitly in terms of $e(4,0)$. Although all the above equations have been solved, for the sake of keeping the discussion here short and to the point, we  restrict the presentation of the results to the  version of these equations where we drop the terms bilinear in covariant tensors. However, to get a feeling for the full theory
we present  what ultimately will be the full spin 3 Cotton (spin 2 covariant) equation
\beq
D_{[\mu}f_{\nu]}{}^{ab}+f_{[\mu}{}^c\tf_{\nu]}{}^{d(a}\ep_{cd}{}^{b)}+f_{[\mu}{}^{(a}\tf_{\nu]}{}^{b)}-\tfrac{1}{3}\eta^{ab}f_{[\mu}{}^{c}\tf_{\nu]}{}^{c}=0,
\eeq
which,  as will become clear below, is a rather complicated equation fifth order in derivatives. We see, for instance, that the spin 2 Schouten tensor $f_{\mu}{}^a$ appears again but now multiplied by  tensors, $\tf_{\mu}{}^a$ and $ \tf_{\mu}{}^{ab}$, from the spin 3 sector. In the simplified discussion below only the first term in this equation is kept, and similar for the other equations. Note, however, that apart from the derivative terms in the transformation rules and field equations also the terms containing the dreibein are kept since these play a crucial role when solving the system of equations.

As just mentioned, solving for $f_{\mu}{}^{ab}(0,4)$ as well as for all the other fields explicitly in terms of $e_{\mu}{}^{ab}(4,0)$ can be done exactly. Our aim now is to 
explain the structure of  the equations and how to solve them. This is best done in the linear approximation (as defined above)  
and using only tensors which are irreps in the flat indices. Then the $F=0$ equations read\footnote{Note that each equation (and field) contains several irreps if all indices are taking into account. E.g., since $F^{ab}(4,0)=0$ is a set of 2-form equations, it contains three equations in the irreps ${\bf 7}+{\bf 5}+{\bf 3}(={\bf 3}\times {\bf 5})$.}
\beqa
\label{F40}
F^{ab}(4,0)&=&0:\,\,\,\pd_{[\mu}e_{\nu]}{}^{ab}+\ep_{[\mu}{}^{c(a}\te_{\nu]c}{}^{b)}=0,\\
\label{F31}
\tF^{ab}(3,1)&=&0:\,\,\,\pd_{[\mu}\te_{\nu]}{}^{ab}-2\ep_{[\mu}{}^{c(a}\tom_{\nu]c}{}^{b)}=0,\\
\label{C1}
\tF^{a}(3,1)&=&0:\,\,\,\tom_{[\mu\nu]}{}^{a}+\tfrac{3}{2}\ep_{[\mu}{}^{ba}\tom_{\nu]b}-2e_{[\mu}{}^a\tb_{\nu]}=0,\\
\label{F22}
\tF^{ab}(2,2)&=&0:\,\,\,\pd_{[\mu}\tom_{\nu]}{}^{ab}+3\ep_{[\mu}{}^{c(a}\tf_{\nu]c}{}^{b)}-e_{[\mu}{}^{(a}\tf_{\nu]}{}^{b)}-\tfrac{1}{3}\eta^{ab}\tf_{[\mu\nu]}=0,\\
\label{C2}
\tF^{a}(2,2)&=&0:\,\,\,\pd_{[\mu}\tom_{\nu]}{}^{a}+3\tf_{[\mu\nu]}{}^{a}-3\ep_{[\mu}{}^{ba}\tf_{\nu]b}=0,\\
\label{C3}
\tF(2,2)&=&0:\,\,\,\pd_{[\mu}\tb_{\nu]}+\tfrac{8}{3}\tf_{[\mu\nu]}=0,\\
\label{F13}
\tF^{ab}(1,3)&=&0:\,\,\,\pd_{[\mu}\tf_{\nu]}{}^{ab}-4\ep_{[\mu}{}^{c(a}f_{\nu]c}{}^{b)}=0,\\
\label{C456}
\tF^{a}(1,3)&=&0:\,\,\,\pd_{[\mu}\tf_{\nu]}{}^{a}+6f_{[\mu\nu]}{}^{a}=0,\\
\label{F04}
F^{ab}(0,4)&=&0:\,\,\,\pd_{[\mu}f_{\nu]}{}^{ab}=0,
\eeqa
which are written in a gauge that does not affect $\tb_{\mu}$ but restricts the  fields (in the spin 3 sector) with one flat index as follows:
\beq
\te_{\mu}{}^a=0,\,\,\,\tom_{\mu}{}^a=e_{\mu}{}^a\tom,\,\,\,\tf_{\mu}{}^a=\ep_{\mu}{}^{ab}\tf_b+e_{\mu}{}^a\tf.
\eeq
These equations  also serve the purpose to define the new fields on the right hand side of the last two equations. Using the irrep ${\bf 3}$ and ${\bf 1}$ parts of the equations 
  $F^{a}(3,1)=0$ to solve for $\tb_{\mu}$ and $\tom$ and similarly the
${\bf 3}$ and ${\bf 1}$ parts of $F^{a}(2,2)=0$ to solve for $\tf_{\mu}$ and $\tf$, there are only fields in the ${\bf 5}$ left to consider, i.e., $f_{\mu}{}^{ab}$ etc.
These latter ones can, however, all  be expressed in terms of $e_{\mu}{}^{ab}$ using the equations in the ${\bf 5}$, i.e., $F^{ab}(4,0)=0$ etc. The equation
$F^{ab}(0,4)=0$ is then the spin 3 Cotton equation. We find that
\beq
\om=\tf=0,\,\,\,\tb_{\mu}=\tfrac{1}{4}\tom_{\nu\mu}{}^{\nu},\,\,\,\,\tf_{\mu}=-\tfrac{1}{2}\tf_{\nu\mu}{}^{\nu},
\eeq
and, as an example, we give the  solution for $\tf_{\mu}{}^{ab}$ obtained from $\tF^{ab}(2,2)$:
\beq
\tf_{\mu}{}^{ab}=-\tfrac{1}{3}(2\ep^{\nu\rho(a}\pd_{\nu}\tom_{\rho\mu}{}^{b)}-\ep_{\mu}{}^{\nu\rho}\pd_{\nu}\tom_{\rho}{}^{ab})+
\tfrac{11}{48}\eta^{ab}\ep^{\nu\rho d}\pd_{\nu}\tom_{\rho d \mu}-\tfrac{1}{48}\ep^{\nu\rho d}e_{\mu}{}^{(a}\pd_{\nu}\tom_{\rho d}{}^{b)}.
\eeq

The equations that are not yet solved at this point are constraints on the only field remaining in the theory\footnote{The spin 2 dreibein is, however, also present since it is used to convert flat to curved indices.} , namely the spin 3  "frame" field  $e_{\mu}{}^{ab}$. These constraints are
\beqa
&&C1:\,\,\,\,\tF^a(3,1)|_{\bf 5}=0\,\,\rightarrow \,\,\ep^{\mu\nu(a}\tom_{\mu\nu}{}^{b)}=0,\\
&&C2:\,\,\,\,\tF^a(2,2)|_{\bf 5}=0\,\,\rightarrow \,\,\ep^{\mu\nu(a}\tf_{\mu\nu}{}^{b)}=0,\\
&&C3:\,\,\,\,\tF^(2,2)|_{\bf 3}=0\,\,\rightarrow \,\,\pd_{[\mu}\tb_{\nu]}+\tfrac{8}{3}\ep_{\mu\nu}{}^a\tf_a=0,\\
&&C4:\,\,\,\,\tF^a(1,3)|_{\bf 5}=0\,\,\rightarrow \,\,6\ep^{\mu\nu(a}f_{\mu\nu}{}^{b)}+\ep^{\mu\nu(a}\pd_{\mu}\tf_{\nu}{}^{b)}=0,\\
&&C5:\,\,\,\,\tF^a(1,3)|_{\bf 3}=0\,\,\rightarrow \,\,\pd_{a}\tf_{\mu}{}^{a}+6f_{\nu\mu}{}^{\nu}=0,\\
&&C6:\,\,\,\,\tF^a(1,3)|_{\bf 1}=0\,\,\rightarrow \,\,\pd_{\mu}\tf^{\mu}=0,
\eeqa
where we can drop the last one as an independent constraint since it follows from $C3$. After some algebra, expressing $\tom_{\mu}{}^{ab}$
in terms of first $\te_{\mu}{}^{ab}$ and then in terms of $e_{\mu}{}^{ab}$ one can verify that $C1$ is an identity. The same result can be seen to 
be true for $C2$. Finally, here we need also to make sure that also $C3$ is identically satisfied. This can be checked as follows. 

We start by noting that when expressed in terms of $\te_{\mu}{}^{ab}$, $C3$ reads :
\beq
\Box\te_{\nu\mu}{}^{\nu}-\pd_{\rho}\pd_{\nu}\te^{\rho\nu}{}_{\mu}-\pd_{\rho}\pd_{\nu}\te_{\mu}{}^{\rho\nu}+\pd_{\mu}\pd_{\nu}\te_{\rho}{}^{\rho\nu}=0.
\eeq
This equation is, however, an identity once the explicit expression of $\te_{\mu}{}^{ab}$ in terms of $e_{\mu}{}^{ab}$ is inserted. Thus, to verify that $C3$ is an identity it seems necessary  to use 
four levels of the above relations, $\tf^{ab}\rightarrow\tom^{ab}\rightarrow\te^{ab}\rightarrow e^{ab}$, which is rather curious.

This shows that the system of equations in the spin 3 case in a rather nice way can be used to express the whole spin 2/spin 3 sector in terms of only two fields, namely
the  dreibein  $e_{\mu}{}^a$ and the basic  spin 3 field $e_{\mu}{}^{ab}$. The corresponding metric tensors are of course the symmetric parts of these fields.  The  gauge symmetries that we have kept intact throughout the analysis are
\beq
\de e_{\mu}{}^a(2,0)=D_{\mu}\Lam^a(2,0)-\ep^a{}_{bc}e_{\mu}{}^b(2,0)\Lam^c(1,1)-e_{\mu}{}^a(2,0)\Lam(1,1),
\eeq
and 
\beqa
\de e_{\mu}{}^{ab}(4,0)&=&D_{\mu}\Lam^{ab}(4,0)+e_{\mu}{}^c(2,0)\tLam^{d(a}(3,1)\ep^{b)}{}_{cd}- (e_{\mu}{}^{(a}(2,0)\tLam^{b)}(3,1)-trace)\cr
&&-\Lam^c(2,0)\te_{\mu}{}^{d(a}(3,1)\ep^{b)}{}_{cd}+(\Lam^{(a}(2,0)\te_{\mu}{}^{b)}(3,1)-trace)\cr
&&-2\Lam(1,1)e_{\mu}{}^{ab}(4,0)-2\Lam^c(1,1)e_{\mu}{}^{d(a}(4,0)\ep_{cd}{}^{b)}.
\eeqa
By using the symmetries with parameters $\Lam^a(1,1)$ and $\tLam^{ab}(3,1)$ we  can  restrict the dreibein fields to their symmetric parts. In fact,  relative flat space the symmetric (but not traceless) part of these fields, denoted as usual $g_{\mu\nu}$ and $g_{\mu\nu\rho}$,  then transform as
\beq
\de g_{\mu\nu}=\pd_{(\mu}\Lam_{\nu)}-\eta_{\mu\nu}\Lam,\,\,\,\de g_{\mu\nu\rho}=\pd_{(\mu}\Lam_{\nu\rho)}-\eta_{(\mu\nu}\tLam_{\rho)},
\eeq
which are  well-known transformation rules; for  the spin 3  generalization  of "diffeomorphisms" and "dilatations", see e.g. \cite{Bergshoeff:2009tb}.

As a final comment in this section we note the following. In the pure spin 2 case there is a well-known on-shell relation between the diffeomorphism parameter $\xi^{\mu}(x)$
and the local translations $\Lam^a(x)(2,0)$ and Lorentz parameters $\Lam^a(x)(1,1)$ in the Chern-Simons formulation \cite{Horne:1988jf}:
\beq
\Lam^a(2,0)=\xi^{\mu}e_{\mu}{}^a,\,\,\,\Lam^a(1,1)=\xi^{\mu}\om_{\mu}{}^a,\,\,\Rightarrow \de_{\xi}=\de_P+\de_M,
\eeq
where the last equation is verified by letting it act on the dreibein and the spin connection and making use of, respectively,  the CS "field equations" 
$F^a(2,0)=0$ and $F^a(1,1)=F(1,1)=0$. Similarly, checking this for the Schouten tensor will require the use of the Cotton equation
.
Of course, this relation between these symmetries should hold also in the higher spin sectors. Checking it on the spin 3 frame field $e_{\mu}{}^{ab}$ shows, e.g.,  that,
now modulo the "field equation" $F^{ab}(4,0)=0$, also
the spin 3 "translation" parameter must be related to $\xi^{\mu}$ as $\Lam^{ab}(4,0)=\xi^{\mu}\,e_{\mu}{}^{ab}$.\footnote{These considerations should be extended to
the relation between the spin 3 generators $\Lam^{ab}(4,0)$, etc and $\xi^{\mu\nu}$ denoted $\Lam^{\mu\nu}$ above.}

\section{Towards an action formulation for all higher spins}

In order to discuss the  Lagrangian we now return to the full non-linear equations based on the Poisson bracket.  We start by noting that there are many ways to present the Lagrangian   depending on how the derivatives are distributed between the fields, e.g.,  in the kinetic terms. Here we will
advocate the use of a particularly symmetric form  that generalizes the standard Chern-Simons action often used for both spin 1 and 2. In the language of this paper, the spin 2 Chern-Simons like action reads 
\beq
\label{CS-action-spin-1}
S_2=\thalf \int (\om_1^a d \om_{1a} + \tfrac{1}{3}\ep_{abc}\om_1^a \wedge \om_1^b \wedge \om_1^c),
\eeq
where  $\om^a_1:=\om^a(1,1)(e)$, that is,  the second order form of the $SO(1,2)$ spin-connection $\om_1=\om^a(1,1)M_a(1,1)$ discussed in section 2. The addition of spin 3 terms to the action will then force us into the star product formulation  as will become clear below. Here the star product will, however, only be used in a perturbative fashion, i.e., by introducing in a stepwise
manner higher and higher multi-commutators
(which can be viewed as coming from an expansion of  $[A,B]_*$) of operators Weyl-ordered  in the  quantized variables $q^{\al},p_{\al}$.

The action for the combined conformal spin 2/spin 3 system is naturally given as a direct generalization of the
one above for spin 2 by making use of $\om(1,1)$ and $\om(2,2)$ expressed in terms of their respective basic spin 2 and spin 3 fields,
$e_{\mu}{}^a$ and $e_{\mu}{}^{ab}$. In the corresponding  $AdS_3/dS_3$ case such an action was given in \cite{Campoleoni:2010zq} although in a first order formalism contrary to 
 the conformal case discussed here. The action presented in  \cite{Campoleoni:2010zq} reads
\beqa
\label{poincare3}
S&=&\tfrac{1}{8\pi G}\int(e^a\wedge (R_a(\om)-2\si \ep_{abc}\om^{bd}\we\om^c{}_d)-2\si e^{ab}\we D\om_{ab}\cr
&&+\tfrac{1}{6l^2}\ep_{abc}(e^a\we e^b\we e^c-12\si e^a\we e^{bd}\we e^c{}_d)),
\eeqa
where the dualized curvature $R_a(\om)=d\om_a+\thalf\ep_{abc}\om^b\wedge \om^c$ and $D\om_{ab}$ is written using the spin 2 Lorentz covariant  derivative $D=d+\om$. The last term puts the theory in $AdS_3$ (with $\Lambda=-\tfrac{1}{l^2}$) and the sign of the parameter $\si$ (either one is allowed by the Jacobi identities)
affects the nature of the gauge group of the Chern-Simons formulation leading to this action: $\si>0$ corresponds to $SU(1,2)\times SU(1,2)$ and $\si<0$ to the more 
familiar case $SL(3,{\bf R})\times SL(3,{\bf R})$.

 As we will now argue there is a corresponding action for the conformal theory of the spin 2/spin 3 part of the higher spin system but  in a second order formalism.  We start this discussion in terms of Poisson brackets.  In fact, the action 
\beq
\label{spin2-3-action}
S=S_2(\om(1,1))+S_3(\tom(2,2)),
\eeq
where the spin 2 term was given above in (\ref{CS-action-spin-1}) and the spin 3 part reads\footnote{There are three irreducible parts in $\tom_{\mu}{}^{ab}(2,2)$ which may in fact appear with different coefficients. In particular $\tb_{\mu}(2,2)=\frac{1}{4}\tom_{\nu\mu}{}^{\nu}(2,2)$ could be present as an additional term $\tb d \tb$.}
\beq
\label{kinetic-spin3}
S_3 = \thalf  \int \tilde \om^{ab}(2,2) D \tilde \om_{ab}(2,2),
\eeq
should capture the (single commutator) Cotton equations for these two values of the spin.
Here the coefficient $\thalf$ gives the spin 3 field a canonically normalized kinetic term but as will become clear below it is also fixed by the interaction term since it contributes  to the field equation for $\om_1$ once the variables $q^{\al},p_{\al}$ are quantized and  the Poisson bracket is replaced by a star product bracket. 
 The covariant derivative $D=d+\om(1,1)$ in (\ref{kinetic-spin3})  
makes the  spin 2 Lorentz symmetry manifest and it would be interesting to see if the other symmetries remaining after gauge fixing (see previous sections) can be used to extend the action
to higher spins.  Note that a 
 cubic term involving three $\tom^{ab}(2,2)$ connections cannot appear.
In order to be consistent the action in (\ref{spin2-3-action}) must, however, be viewed as written in a star product formulation as we now explain.

From the algebraic construction of the interaction terms in the previous section 
we  see that  since the kinetic term  $\tom^{ab}(2,2)d\,\tom_{ab}(2,2)$ is in level  $(4,4)$ any interaction term emerging from a single commutator (i.e., the Poisson bracket formulation) will contain fields which together correspond to level  $(5,5)$,
i.e., two $\tom^{ab}(2,2)$ fields and one $\om^{a}$(1,1). Thus
 the interaction  term is $\tom^{ab}(2,2) \ep_{acd}\om^c (1,1)\tom_b{}^{d}(2,2) $ which then completes the covariant derivative $D$ in (\ref{kinetic-spin3}). One should note, however,  that the presence of this cubic term in the  action (\ref{kinetic-spin3}) implies a contribution to  the spin 3 Cotton equation which is a single commutator effect (that is, it is part of the Poisson bracket formulation) but  its effect  in the spin 2 
 Cotton equation is a triple commutator term whose origin is in the star product formulation. Indeed, the spin 2 equation derived in the previous section did not contain such a term.  That it does come from a triple commutator has been verified explicitly.
 
 A direct consequence of the facts just described is that the field equations obtained from solving the algebraic equations in the Poisson bracket formulation of the higher spin theory expressed in terms of the equation $F=0$ are not integrable, i.e., they are not compatible with
 a Lagrangian.  Adding the terms needed to make the theory compatible with a Lagrangian seems to be equivalent to introducing the multi-commutators that turns the Poisson bracket into
 a full star product bracket $[A,B]_*$.
 
 Another way to state this conclusion is as follows: the whole star product Lagrangian  seems to be reproduced by writing down a Lagrangian that captures the Poisson bracket sector  of the field equations for all spins in the theory. This follows since all  star product interaction terms in the action are related to a single commutator term  in a field equation for some sufficiently high spin.  One may consider, for instance, terms of the form 
 $\tom^{a_1...a_p}(p,p) \ep_{a_1bc}\om^b (1,1)\tom^c{}_{a_2...a_p}(p,p) $. They will appear in the spin 2 field equation for  positive integer values of $p$ as a {\it (2p-1)-commutator} effect while they  come from a single commutator in the {\it spin p+1} field equations for $\tom^{a_1...a_p}(p,p)$. These are higher spin generalizations of $\om^a(1,1)$ and appear in the generator decomposition at level spin $s=n+1$:
 \beqa
\label{spin-s-sequence}
&&f^{a_1a_2...a_n}(0,2n),\,\tf^{a_1a_2...a_n}(1,2n-1),\tf^{a_1a_2...a_{n-1}}(1,2n-1),.... \cr
&&\tom^{a_1a_2...a_n}(n,n),\,\tom^{a_1a_2...a_{n-1}}(n,n),...,\tom(n,n),\cr
&&.....,\te^{a_1...a_n}(2n-1,1),  \te^{a_1...a_{n-1}}(2n-1,1),\,e^{a_1...a_n}(2n,0). 
\eeqa
  
 We thus conclude that the following Lagrangian is consistent with the star product in the above sense:
 \beq
 \label{spin-2-3}
 S_{spin-2-3}=Tr\int (\om_1d\om_1+\tfrac{1}{3}\om_1^3)+\thalf Tr\int(\tom_2D\tom_2).
 \eeq
 Here $\tom_n=\tom_{a_1...a_n}\tM^{a_1...a_n}$ (but without tilde for $n=1$) and  the kinetic terms are normalized in a canonical way. We have  also introduced 
the trace $Tr(q^{\al}p_{\be})=\de^{\al}_{\be}$ which can be generalized to the entire higher spin algebra.  The verification of the coefficients appearing in (\ref{spin-2-3})  involves a cross-check (the term $\om_1\tom_2\tom_2$) giving at least some support to the idea that this can be generalized to all higher spins.
 
  To summarize the above discussion we have seen that as higher commutator terms in the star product are taken into account   new interaction terms involving  higher spin fields appear.    This is clear since as soon as we 
change to quantum operators  multi-commutators will arise
which reduce the $(a,b)$ level of the interaction term by $(1,1)$ for each new commutator that is computed. The star product  will thus induce
new  contributions from (almost) all higher spins to any  lower spin  field equation.
We now also understand that these higher commutator terms are necessary for a Lagrangian formulation to exist. 

The above observations suggest that the structure of the action for spin 2 and 3 may be generalized to any spin. By using the trace introduced above
 we obtain  unique coefficients for all higher spin terms 
$S_{s=n+1} \propto \int \tilde \om_{a_1...a_n}(n,n) d \tilde \om^{a_a...a_n}(n,n)$ in the action. The all-spin $(s\geq 2)$ kinetic part of the action may  then be 
written
\beq
\label{kinetic-allspinaction}
S= \int Tr( \tilde \Om d \tilde \Om) ,
\eeq
where we have defined the 1-form $\tOm=\Sigma_{n=1}^{\infty}a_n\tom^{a_1..a_n}\tM_{a_1...a_n}$ where all fields $\tom^{a_1..a_n}$ are assumed to be expressed in terms of the corresponding
higher spin  1-forms $e^{a_1..a_n}$. The assumption made here is thus that the spin $s=n+1$  "connection" 1-form $\tom^{a_1..a_n}$, 
which is  symmetric and traceless in $a_1...a_n$, can be expressed in terms of the spin $s $ hyper-dreibein (or frame field) $e_{\mu}{}^{a_1...a_n}$ and that all other fields 
can be solved for and the constraints shown to be identities as done for spin 2 and 3 above. This discussion is an explicit realization of the application of 
higher spin $\si_-$ cohomology to the conformal theory in three dimensions as described in \cite{Vasiliev:2012vf}\footnote{This analysis may be compared to one of the four-dimensional super-higher spin theory carried out in \cite{Sezgin:1998gg, Sezgin:2002ru}, see also \cite{Vasiliev:1999ba} and referencies therein.
The fields of the $AdS$ Vasiliev-type theory discussed there are divided into those that can be set to zero by gauge transformations, those that can be solved for and the rest which constitute the physical fields. However, the field equations are quite different from the conformal ones we are dealing with here and those systems do not seem to contain the same type of constraints as in our case.}. For some properties of the trace relevant in our context of the star product, see, e.g., 
\cite{Bekaert:2010ky}.

Adding the interaction terms to the kinetic all-spin action in (\ref{kinetic-allspinaction})  is an intricate problem.  
%
%
This action  captures the whole kinetic term part of the  higher spin theory based on the Poisson bracket and, for reasons given above, also a number of star product terms when it is made spin 2 Lorentz covariant. We can then use it as the starting point for 
the construction of a consistent Lagrangian theory coming from the star product  by adding the effects of the remaining interaction terms that arise in the Poisson bracket theory: 
starting from spin four the field equations will contain new interaction terms not involving $\om_1$ giving new terms in the action. The terms we have in mind here are terms  like the last one in $S_4=\int(\tom_3d\tom_3+\tom_3\om_1\tom_3+\tom_3\tom_2\tom_2)$ where  $\tom_3$ refers to the top irrep of $\om(3,3)$ etc. Two of the main constraining factors are now the dimensionality of the tensors used to construct the interaction terms (the 1-form $\tom_n$ is dimensionless for all $n$) and the trace properties of the three sets of flat indices $a_1a_2....a_n$   (being symmetric and traceless).

 An action for all spins that  contains the terms mentioned above  may then be 
 written\footnote{The introduction of the coefficients $a_n$  in $\tOm$ means that the few terms discussed in this paper do not provide any non-trivial cross checks on this form of the action.}
\beq
\label{allspinactionwithcubeterms}
S=  \int Tr( \tilde \Om d \tilde \Om+\tfrac{2}{3}\tOm\wedge \tOm \wedge \tOm),
\eeq
where the multi-commutator formulation used here corresponds to a star product although this is not explicitly indicated by the 
notation\footnote{The cubic terms in this star product action are quite restricted: only terms $\tom_m\wedge \tom_n \wedge \tom_p$ with $m+n+p=$ {\it odd integer} and $m+n\geq p+1$ and cyclic can occur. This means that each such star product term will always contain a one-commutator contribution to the action which  provides a kind of  Poisson bracket limit of the star product theory.}. 
The coefficients in (\ref{allspinactionwithcubeterms}) have  been determined by checking a small number of terms
and it would be interesting to see if a more general set of terms  follow the  same pattern. One may note the similarity  between the  field equations derivable from this all-spin action and
the set of field 
equations for $\tom_n$ in the star product equation $F=0$ although in reality they are extremely different. 

It would be rather remarkable if the action (\ref{allspinactionwithcubeterms}) captures the complete conformal higher spin theory.  The main assumption is then that once the kinetic terms are written down in terms of the HS spin connections (which can always be done) the interaction terms  do not need any of the other fields in the spin $s=n+1$ sequences
(\ref{spin-s-sequence})
 in order to describe the entire star product higher spin theory. However, should that not be the case then
 (\ref{allspinactionwithcubeterms}) may correspond to a subsector\footnote{We are not assuming that this subsector constitutes a consistent truncation of the whole theory.} of the full Lagrangian which  can be completed by reanalyzing the field equations coming from 
solving $F=0$ after including higher terms in the star product and comparing them to the field equations from the action proposed here. Clearly such terms are present, as for instance  the triple commutator term in the following $M_a$ equation of the spin 2 Cotton system 
\beq
\label{lorentzspintwo}
F^L_a(1,1)=d\om_a(1,1)+\thalf \{\om(1,1),\om(1,1)\}|_{M_a}+\thalf \{\om(2,2),\om(2,2)\}^{(3)}|_{M_a}+...=0,
\eeq 
where terms not involving explicit spin connections can also occur, e.g., $\te(n,1)\tf(1,n)|_{M^a}$. However, if this actually means that such terms should be added also to the HS  action in 
(\ref{allspinactionwithcubeterms}) is not clear\footnote{Recall that (\ref{lorentzspintwo}) is in the first order formulation while (\ref{allspinactionwithcubeterms}) is in the second order one.}.

A similar strategy can be adopted when we now turn to the problem of coupling this theory for spin 2, 3, 4... to matter in the form of scalars. As we will see below the classical formulation in terms of Poisson brackets leads  immediately to problems and must be  amended by star product corrections. Here we will find that instead of the spin connections $\tom(n,n)$ it is the fields in the level just above them, namely $\tf(n-1,n+1)$ in the spin one representation, that appear naturally. 

\section{The coupling of higher spins to scalar matter}

Compared to the discussion in the previous section the coupling to scalar and/or spinor fields is an entirely different problem which requires unfolding techniques well-known from Vasiliev-type constructions
of interacting higher spin systems in $AdS$ as well as from  higher spin theories with conformal  symmetry \cite{Vasiliev:1992gr,Shaynkman:2004vu}. Unfolding in the context of the above Poisson bracket formulation is straightforward\footnote{One has, however, to define the dilatation operator to have a specific eigenvalue on the representation used for the scalar field. The fact that quantization is the obvious way to obtain this eigenvalue indicates that one must
eventually take the whole star product into account as in Vasiliev-type constructions. In the context of W-algebras  quantization was  performed
 in \cite{Pope:1990kc}, see also \cite{Bergshoeff:1991un,Pope:1991uz}. For a more recent discussion in three dimensions, see \cite{Ammon:2011ua}.}  and leads (almost) directly to field equations for the
matter fields that are (spin 2) covariant under the higher spin algebra in the sense of the previous section. However, imposing $D\Phi(x;p,q)=0$ does not produce the conformal coupling term appearing in the conformal scalar field equation $\Box\phi(x)-\tfrac{1}{8}R\phi=0$. To retrieve this equation one has to in effect quantize the $p,q$ system, impose the equation $D\Phi(x;p)|0>=0$  and keep only the single commutator contributions. Note that the geometric structure of the higher spin theory that emerged above when solving $F=0$ based on  the Poisson bracket  does not change at all if one instead uses this approximation of the quantized version\footnote{ E.g., all factors of "$i$" cancel out which requires adopting the expansion $A=-ie_aP^a+...$ as familiar from the star product formulation.}. 

Adding the spin 3 sector to the scalar field equation 
gives the equation\footnote{For unfolding to work the untraced version of this equation  must contain the symmetric and traceless 
field $\phi_{\mu\nu}$ in the master field $\Phi$. This is easily seen to be the case.}
\beq
\label{scalar-spin3-eq}
\Box\phi(x)-\tfrac{1}{8}R\phi(x)+\tf\phi(x)=0,
\eeq
where now $\tf:=\tf_{\mu}{}^{\mu}(1,3)$ has appeared in analogy with the second term that comes  from the trace of the Schouten tensor $f_{\mu}{}^{a}(0,2)$. Note that it is not the spin 3  tensor $f_{\mu}{}^{ab}(0,4)$ (which perhaps should be called the spin 3 Schouten tensor) that appears here but instead
a tensor from the level below: $\tf_{\mu}{}^a(1,3)$. It has the correct dimension and a non-zero trace in the full non-linear solution of the system. From the complete analysis\footnote{Complete expressions for these tensor fields are easily obtained and will be presented elsewhere.} we know that
\beq
\tf(1,3)=\tf_{\mu}{}^{\mu}(1,3)=-\thalf\ep^{\mu\nu\rho}\te_{\mu\nu b}(3,1)f_{\rho}{}^b(0,2),
\eeq
where on the right hand side we recognize the spin 2 Schouten tensor and the spin 3 field $\te_{\mu}{}^{ab}(1,3)$ which is expressed in terms of the basic spin 3 field $e_{\mu}{}^{ab}(0,4)$ as follows\footnote{Note that in a metric gauge where the basic fields are totally symmetric, i.e., $e_{\mu}{}^{ab}$ is replaced by $g_{\mu ab}$, the field $\te_{\mu}{}^{ab}$ is not getting symmetrized and hence $\tf$ remains non-zero.}
\beq
\te^{abc}(3,1)=\ep^{\mu\nu a}D_{\mu}e_{\nu}{}^{bc}(4,0)-2(\ep^{\mu\nu (b}D_{\mu}e_{\nu}{}^{c)a}(4,0)-\tfrac{1}{3}\eta^{bc}\ep^{\mu\nu}{}_dD_{\mu}e_{\nu}{}^{da}(4,0)).
\eeq
The new spin 3 term in (\ref{scalar-spin3-eq}) has its origin in  the fact that the Weyl 
ordered form of the generator $\tK^a(1,3)$ gives $K^a(0,2)$ when acting on the vacuum state, a fact that should generalize directly to higher spins.

At this point one has to address the issue of back reaction  that  is solved for  theories in $AdS$ by Vasiliev's construction of the interaction terms in the field 
strength equation (see, i.e., \cite{Vasiliev:1995dn,Vasiliev:1999ba} and references therein). However, this kind of  solution  is  not available in this conformal theory (see, e.g., the  discussion in \cite{Vasiliev:2012vf}). In the non-linear and spin 2 covariant  formulation used  here 
the action will contain the following 
terms for spin 2 and 3 coupled to the scalar field:
\beq
S=S_2(\om(1,1))+S_3(\tom(2,2))+\int d^3x\,e(-\thalf g^{\mu\nu}\pd_{\mu}\phi\pd_{\nu}\phi-\tfrac{1}{16}R\phi^2+\thalf \tf\phi^2).
\eeq
There is, of course, no new insight provided by writing down this part of the action unless it helps in generalizing it to higher spins. One immediate observation is, however,  that the field(s) in the general spin $s$ sequence (\ref{spin-s-sequence}) that will appear in this action coupled to $\phi^2$ belong(s) to the level just above the corresponding spin connection. To gain further insight one  could try to implement the higher spin symmetries that are still manifest after gauge fixing 
or use covariant Noether techniques of the kind successfully employed recently in a similar context namely
topologically gauged $M2$ brane-like matter systems \cite{Gran:2008qx,Chu:2009gi,Nilsson:2013fya,Chu:2010fk,Gran:2012mg,Nilsson:2012ky}.
Alternatively one could  just compute further terms in the spirit of the previous sections. If any of these  approaches will provide a fruitful way forward  remains to be seen, however. Even more challenging is perhaps the issue
of consistency of the star product theory including the scalar field. In fact, it seems not to be known how to construct  such an interacting theory from first principles. 

\section{Conclusions}

In this paper we aim at  developing a fully non-linear spin 2 covariant formulation of a $CFT_3$ based on a higher spin algebra that approximates the star product commutator by the first term, the Poisson bracket or the single commutator after quantization.  
One can then hope to in a systematic way introduce the effects of multi-commutators in the Lagrangian and retrieve the full star product formulation. 

To this end we propose to  start from an action for the kinetic terms constructed in terms of the generalized spin $s=n+1$ connections $\tom_n:=\tom^{a_1...a_n}(n,n)$ which are symmetric and traceless in the indices and expressed 
solely in terms of the corresponding frame fields $e^{a_1...a_n}$. This can always be done since it just amounts to a symmetric distribution of the derivatives in the case of the kinetic term. The strategy is then  to add  to the action the interaction terms generated by the Poisson higher spin algebra 
thereby also introducing a set of multi-commutator terms into the Lagrangian formulation of the theory. 
As discussed previously in this paper, for each spin $s$ field equation the new interaction terms involve fields with higher and higher spin  for each new multi-commutator that is taken into account. The final form of the field equations are then written in  terms of star products which is an integrable set of equations, i.e., they are compatible with a Lagrangian formulation which is not the case for the field equations in the Poisson bracket formulation. Specific  examples that  support this picture are discussed in detail.  Also the coupling to scalars can be obtained in a consistent fashion including back-reaction on the higher spins through the use of the Lagrangian as discussed in the previous section. In these scalar field coupling terms we found that it is not the spin connections themselves that appear naturally but rather fields one level higher
in the spin sequences (\ref{spin-s-sequence}), namely for each spin $s=n+1$ one of the fields denoted $\tf(n-1,n+1)$.

The virtue of this procedure is  that it  introduces couplings among higher spin  fields expressed in terms  of their associated spin 2 covariant tensors and that interaction terms of particular interest may be constructed without deriving the whole Lagrangian. This way sums over all spins
may be obtainable for certain kinds of terms and used to derive  all-spin effects like, e.g., the total Weyl anomaly that has been computed in recent works by Giombi et al  \cite{Giombi:2013yva} and by Tseytlin \cite{Tseytlin:2013jya}. It is an interesting fact that
these anomalies turn out to cancel in certain cases (dimensions).  In the context of W-algebras in two dimensions such cancelations were  found already in \cite{Bergshoeff:1991un},
see also \cite{Pope:1991uz, Pope:1990vf}.

There are a number of key issues that must be addressed before the usefulness of this approach can be assessed. The most crucial ones are to check whether other fields
than the $\tom(e)$:s appear in the interaction terms  and if the coupling to scalar fields can be set up in a way that produces a completely consistent star product  theory in the end.
Assuming that this conformal higher spin theory coupled to scalars does exist, implementing supersymmetry in a manner similar to \cite{Gran:2008qx,Chu:2009gi,Nilsson:2013fya,Chu:2010fk,Gran:2012mg,Nilsson:2012ky} for the topologically gauged (i.e., coupled to superconformal gravity) models would then be an interesting next step
\footnote{Note that the Chern-Simons formalism used in this paper does not eliminate the possibility to add a source term on the RHS of the Cotton equation. Indeed, the Bianchi identities
are satisfied also if one sets to zero all components of  $F=0$ except the last one  corresponding to the Cotton equation.}.   If the new $\phi^6$ interaction  terms for the scalar fields  found in these papers survive the coupling to higher spins than 2 then the breaking of conformal  symmetries found  in \cite{Chu:2009gi} and discussed in detail  in \cite{Nilsson:2013fya} (see also \cite{Chu:2010fk,Gran:2012mg}) 
might have interesting implications also for the higher spins.

Another important issue   is the connection of a conformal higher spin theory in three dimensions to some Vasiliev type theory on $AdS_4$.
To produce the conformal higher spin theory studied here as a boundary theory one has to impose Neumann boundary conditions on gauge fields in $AdS_4$ as emphasized for 
spin 2 in  \cite{Nilsson:2012ky}
and for higher spins in \cite{Giombi:2013fka} (and somewhat indirectly in \cite{Nilsson:2012ky}).  In \cite{Vasiliev:2012vf} Vasiliev actually demonstrates  that  the $AdS/CFT$ correspondence\footnote{$AdS/CFT$ related observations in a higher spin context (without unfolding) were made already in \cite{Bergshoeff:1988jm,Bergshoeff:1988jx}.}  is a natural consequence of unfolding. Thus we would like to express the $SO(2,3)$ higher spin algebra   discussed in this paper and here expressed  in terms of  variables relevant for $CFT_3$ (i.e., the $q^{\al},p_{\al}$)   also
  in terms of variables relevant for the symmetries of  $AdS_4$.
This issue is  discussed by Vasiliev in \cite{Vasiliev:2012vf}. Here we give a slightly different angle on this connection by relating both cases to the unitary realization of the algebra as follows. The variables of the unitary  realization are $a_i,a_i^{\dagger}=(a_i)^{\dagger}$ and satisfy $[a_i,a_j^{\dagger}]=\delta_{ij}$. Their relation to the hermitian $CFT_3$ variables 
$q^{\al},p_{\al}$, which transform in the two-dimensional representation of $sl(2,{\bf R})$ and satisfy $[q_{\al},p_{\be}]=i\ep_{\al \be}$ after lowering the index on $q^{\al}$, is 
\cite{Shaynkman:2001ip}
\beq
a_1=\tfrac{1}{\sqrt{2}}(q_1+ip_2),\,\,\,a_2=\tfrac{1}{\sqrt{2}}(q_2-ip_1),
\eeq
while the $sl(2,{\bf C})$ variable  $y_{\al},\bar y_{\dot \al}=(y_{\al})^{\dagger}$ used in $AdS_4$ are related to $a_i,a_i^{\dagger}$ by \cite{Engquist:2002vr}
\beq
y_1=a_1+ia_2^{\dagger},\,\,\,y_2= -a_2+ia_1^{\dagger},
\eeq
which  satisfy $[y_{\al},y_{\be}]=2i\ep_{\al\be}$.
This way a direct and basically unique connection between the $AdS_4$ and $CFT_3$ descriptions of the higher spin symmetry is established. This should be equivalent to the 
one used in section 7 of \cite{Vasiliev:2012vf}
although it does not look entirely identical.

\acknowledgments

I am very grateful to Ergin Sezgin  and Per Sundell for several constructive comments concerning  the above presentation.

\appendix

\section{Conventions}

\subsection{Three-dimensional gamma matrix relations}
The full higher spin algebra is very easily derived using the spinorial variables $q^{\al}$ and $p_{\al}$,
see for instance the linearized analysis in \cite{Pope:1989vj}. In this paper we have chosen to work with
vector indices instead which implies that  a number of three-dimensional gamma matrix identities
are needed. These are

\beq
(\ga^a)_{(\al\be}(\ga_a)_{\ga)}{}^{\de}=0
\eeq
\beq
(\ga^{[a})_{\al\be}(\ga^{b]})^{\ga\de}= \ep^{ab}{}_c(\ga^{c})_{(\al}{}^{(\ga}\de_{\be)}^{\de)}
\eeq
\beq
(\ga^{[a})_{(\al\be}(\ga^{b]})_{\ga)}{}^{\de}=-\thalf \ep^{ab}{}_c(\ga^c)_{(\al\be}\de_{\ga)}^{\de}
\eeq
\beq
(\ga^a)_{\al\be}(\ga_a)^{\ga\de}=2\de^{(\ga\de)}_{\al\be}=2(\ga^a)_{(\al}{}^{(\ga}(\ga_a)_{\be)}{}^{\de)}
\eeq
\beq
(\ga^a)_{(\al}{}^{(\ga}(\ga^b)_{\be)}{}^{\de)}=\eta^{ab}\de^{(\ga\de)}_{\al\be}-(\ga^{(a}_{\al\be}(\ga^{b)})^{\ga\de}
\eeq
\beq
(\ga^a)_{(\al(\ga}(\ga^b)_{\be)\de)}=\eta^{ab}\thalf(\ep_{\al\ga}\ep_{\be\de}+\ep_{\al\de}\ep_{\be\ga})-\ga^{(a}_{\al\be}\ga^{b)}_{\ga\de}
\eeq

\subsection{The spin 2 - spin 3 Poisson bracket algebra}
The commutators involving the generators in the spin 2 and 3 sectors used in this paper are tabulated below.
Also some commutators involving spin 4 have been used but these are not given here.

The algebra generated by the spin 2 generators $P^a(2,0)$, $M^a(1,1)$, $D(1,1)$ and  $K^a(0,2)$  is
\beqa
[M^a,M^b]&=&\ep^{ab}{}_cM^c,\cr
[M^a,P^b]&=&\ep^{ab}{}_cP^c,\cr
[M^a,K^b]&=&\ep^{ab}{}_cK^c,\cr
[P^a,K^b]&=&-2\ep^{ab}{}_cM^c-2\eta^{ab}D,\cr
[D, P^a]&=&P^a,\cr
[D, K^a]&=&-K^a,
\eeqa
where we have simplified the notation by dropping the $(p,q)$. 

The commutators one generator from each of  the spin 2 and spin 3 sectors
are:
\beqa
[P^a(2,0),\tP^{bc}(3,1)]&=&\ep^{a(b}{}_dP^{c)d}(4,0),\cr
[P^a(2,0),\tP^b(3,1)]&=&-P^{ab}(4,0),\cr
[P^a(2,0),\tilde M^{bc}(2,2)]&=&-2\ep^{a(b}{}_d \tilde P^{c)d}(3,1)-\eta^{a(b}\tilde P^{c)}(3,1)+\tfrac{1}{3}\eta^{bc}\tilde P^a(3,1),\cr
[P^a(2,0),\tilde M^b(2,2)]&=&-\tilde P^{ab}(3,1)+\tfrac{3}{2}\ep^{ab}{}_c\tilde P^c(3,1),\cr
[P^a(2,0),\tilde D(2,2)]&=&-2\tilde P^a(3,1),\cr
[P^a(2,0),\tK^{bc}(1,3)]&=&3\ep^{a(b}{}_d\tilde M^{c)d}(2,2)-3\eta^{a(b}\tilde M^{c)}(2,2)+\eta^{bc}\tilde M^a(2,2),\cr
[P^a(2,0),\tK^{b}(1,3)]&=&-\tM^{ab}(2,2)-3\ep^{ab}{}_c\tilde M^c(2,2)-\tfrac{8}{3}\eta^{ab}\tilde D(2,2),\cr
[P^a(2,0),K^{bc}(0,4)]&=&-4\ep^{a(b}{}_d\tK^{c)d}(1,3)-6\eta^{a(b}\tK^{c)}(1,3)+2\eta^{bc}\tK^a(1,3),\cr
[M^a(1,1),P^{bc}(4,0)]&=&2\ep^{a(b}{}_dP^{c)d}(4,0),\cr
[M^a(1,1),\tP^{bc}(3,1)&=&2\ep^{a(b}{}_d\tP^{c)d}(3,1),\cr
[M^a(1,1),\tP^b(3,1)]&=&\ep^{ab}{}_c\tP^c(3,1),\cr
[M^a(1,1),\tilde M^{bc}(2,2)]&=&2\ep^{a(b}{}_d \tilde M^{c)d}(2,2),\cr
[M^a(1,1),\tilde M^b(2,2)]&=&\ep^{ab}{}_c\tilde M^c(2,2),\cr
[M^a(1,1),\tK^{bc}(1,3)&=&2\ep^{a(b}{}_d\tK^{c)d}(1,3),\cr
[M^a(1,1),\tK^{b}(1,3)&=&\ep^{ab}{}_c\tK^{c}(1,3),\cr
[M^a(1,1),K^{bc}(0,4)&=&2\ep^{a(b}{}_dK^{c)d}(0,4),\cr
[D(1,1),P^{bc}(4,0)]&=&2P^{bc}(4,0),\cr
[D(1,1),\tP^{bc}(3,1)]&=&\tP^{bc}(3,1),\cr
[D(1,1),\tP^{b}(3,1)]&=&\tP^{b}(3,1),\cr
[D(1,1),\tK^{bc}(1,3)]&=&-\tK^{bc}(1,3),\cr
[D(1,1),\tK^{b}(1,3)]&=&-\tK^{b}(1,3)\cr
[D(1,1),K^{bc}(0,4)]&=&-2K^{bc}(0,4),\cr
[K^a(0,2),P^{bc}(4,0)&=&-4\ep^{a(b}{}_d\tilde P^{c)d}(3,1)+6\eta^{a(b}\tilde P^{c)}(3,1)-2\eta^{bc}\tilde P^a(3,1),\cr
[K^a(0,2),\tP^{bc}(3,1)]&=&3\ep^{a(b}{}_d\tilde M^{c)d}(2,2)+3\eta^{a(b}\tilde M^{c)}(2,2)-\eta^{bc}\tilde M^a(2,2),\cr
[K^a(0,2),\tP^b(3,1)]&=&\tM^{ab}(2,2)-3\ep^{ab}{}_c\tilde M^c(2,2)+\tfrac{8}{3}\eta^{ab}\tilde D(2,2),\cr
[K^a(0,2),\tM^{bc}(2,2)]&=&-2\ep^{a(b}{}_d\tilde K^{c)d}(1,3)+\eta^{a(b}\tilde K^{c)}(1,3)-\tfrac{1}{3}\eta^{bc}\tilde K^a(1,3),\cr
[K^a(0,2),\tM^b(2,2)]&=&\tilde K^{ab}(1,3)+\tfrac{3}{2}\ep^{ab}{}_c \tilde K^c(1,3),\cr
[K^a(0,2),\tD(2,2)]&=&2\tilde K^a(1,3),\cr
[K^a(0,2),\tK^{bc}(1,3)]&=&\ep^{a(b}{}_dK^{c)d}(0,4),\cr
[K^a(0,2),\tK^{b}(1,3)]&=&K^{ab}(0,4).
\eeqa

\end{document}